\documentclass[12pt,preprint]{aastex}

\shorttitle{Structure, Propagation and Expansion of a CME-Driven Shock}

\shortauthors{Liu et al.}

\begin{document}

\title{Structure, Propagation and Expansion of a CME-Driven Shock in the Heliosphere:
A Revisit of the 2012 July 23 Extreme Storm} 

\author{Ying D. Liu\altaffilmark{1,2}, Huidong Hu\altaffilmark{1,2}, Bei Zhu\altaffilmark{1,2}, 
Janet G. Luhmann\altaffilmark{3}, and Angelos Vourlidas\altaffilmark{4}} 

\altaffiltext{1}{State Key Laboratory of Space Weather, National Space 
Science Center, Chinese Academy of Sciences, Beijing 100190, China;
liuxying@spaceweather.ac.cn}

\altaffiltext{2}{University of Chinese Academy of Sciences, Beijing 100049, China}

\altaffiltext{3}{Space Sciences Laboratory, University of California, Berkeley, 
CA 94720, USA}

\altaffiltext{4}{The Johns Hopkins University Applied Physics Laboratory, 
Laurel, MD 20732, USA} 

\begin{abstract}

We examine the structure, propagation and expansion of the shock associated with the 2012 July 23 extreme coronal mass ejection (CME). Characteristics of the shock determined from multi-point imaging observations are compared to in situ measurements at different locations and a complex radio type II burst, which according to our definition has multiple branches that may not all be fundamental-harmonic related. The white-light shock signature can be modeled reasonably well by a spherical structure and was expanding backward even on the opposite side of the Sun. The expansion of the shock, which was roughly self-similar after the first $\sim$1.5 hours from launch, largely dominated over the translation of the shock center for the time period of interest. Our study also suggests a bow-shock morphology around the nose at later times due to the outward motion in combination with the expansion of the ejecta. The shock decayed and failed to reach Mercury in the backward direction and STEREO B and Venus in the lateral directions, as indicated by the imaging and in situ observations. The shock in the nose direction, however, may persist to the far outer heliosphere, with predicted impact on Dawn around 06 UT on July 25 and on Jupiter around 23:30 UT on July 27 by an MHD model. The type II burst shows properties generally consistent with the spatial/temporal variations of the shock deduced from imaging and in situ observations. In particular, the low-frequency bands agree well with the in situ measurements of a very low density ahead of the shock at STEREO A. 

\end{abstract}


\section{Introduction}

Coronal mass ejections (CMEs) are large-scale magnetic structures expelled from the Sun. They can drive shocks in the corona and interplanetary space, which are key generators of energetic particles. CME-driven shocks can also significantly compress and distort the magnetosphere, enhancing the geo-effective potential associated with CMEs. Measuring the three-dimensional structure of these shocks and tracking their propagation and expansion in the heliosphere are thus of critical importance for solar and heliospheric studies as well as space weather. 

In situ measurements at multiple spacecraft, although sporadically distributed in the heliosphere, have revealed a large angular extent of CME-driven shocks. \citet{reisenfeld03} report an interplanetary shock from 2001 November, as well as the CME driver, observed at both near-Earth spacecraft and Ulysses with a latitudinal separation of 73$^{\circ}$ and longitudinal separation of 64$^{\circ}$. \citet{liu08} present another CME-driven shock observed at both the Earth and Ulysses in 2006 December when they were separated by 74$^{\circ}$ in latitude and 117$^{\circ}$ in longitude. These two shocks span at least from the ecliptic to a heliospheric pole. The large extent of the shocks indicates that the global configuration of the heliosphere can be altered by a single shock as it sweeps through the solar wind. \citet{liu08} use an MHD model, which assumes spherical symmetry, to connect the in situ measurements of the 2006 December shock at different spacecraft. They suggest that the shock surface on a global scale is more or less spherical (with the center at the Sun) based on the successful model-data comparison. It should be stressed that the fast ambient solar wind in the southern heliospheric pole, where the shock flank was observed, probably had helped to shape the shock (by reducing the deceleration of the flank so that the flank could keep the same pace with the nose).    

CME-driven shocks have been imaged in white light by coronagraphs and heliospheric imagers. They are usually recognized as a faint edge ahead of the CME front in coronagraph observations near the Sun, showing various morphologies such as a bow wave, double front and spheroidal shape \citep[e.g.,][]{vourlidas03, vourlidas09, liu08, liu13, ontiveros09, hess14, kwon14, kwon15}. The bow-shock morphology is observed around narrow CMEs, while the double-front and spheroidal morphologies are associated with wide and fast events \citep{vourlidas09}. The double-front shock structure probably forms when the CME propagates against the pre-existing heliospheric plasma sheet \citep[e.g.,][]{liu09b, liu11}: CME propagation and expansion are restricted along the direction of the heliospheric plasma sheet whereas the CME still expands on both sides, which will give rise to a dent in the CME front as well as the shock. CME-driven shocks in interplanetary space appear as a broad front in heliospheric imaging observations \citep{liu11, liu12, liu13, maloney11, volpes15}. The shock and sheath become a dominant structure in those images, as the density within the CME driver decreases due to expansion in the interplanetary medium. Wide-angle heliospheric imagers such as those from the Solar Terrestrial Relations Observatory \citep[STEREO;][]{kaiser08} enable connections between imaging and in situ observations of CME-driven shocks, which leads to a possible prediction of the shock parameters at 1 AU \citep[e.g.,][]{liu11, liu12, liu13, volpes15}. 

CME-driven shocks can also be tracked in the heliosphere using type II radio bursts, which are well-known shock signatures. They are plasma radiation near the local plasma frequency and/or second harmonic and typically drift downward in frequency \citep[e.g.,][]{nelson85, cane87}. The frequency drift arises from the decrease of the ambient plasma density when the shock moves away from the Sun, so it can be used to characterize shock propagation. Shock distances inverted from type II bursts have been compared with coronagraph observations near the Sun \citep[e.g.,][]{reiner07, liu08, liu09a} and more recently with wide-angle heliospheric imaging data from STEREO over essentially the entire Sun-Earth distance \citep{liu13}. In addition to the distances, the radio observations may also give clues on type II radio source locations with respect to the CMEs in combination with imaging data \citep[e.g.,][]{juan12, feng12, hu16}. In general, they are not widely used as the images because not much information on the shock structure can be inferred from type II bursts.    

The 2012 July 23 extreme CME, which was imaged by three widely separated spacecraft, provides a great opportunity to study the propagation, expansion and structure of CME-driven shocks. This event has attracted significant attention owing to its unusually high solar wind speed and extremely strong ejecta magnetic field at STEREO A \citep[e.g.,][]{russell13, baker13, ngwira13, liu14, temmer15, cash15, riley16, zhang16, zhu16}. \citet{russell13} examine the properties of the shock measured in situ at STEREO A and suggest that the shock was modulated by energetic particles. \citet{baker13}, \citet{ngwira13} and \citet{zhang16} consider the possible geo-effects of the event if it had hit the Earth. \citet{liu14} analyze multi-point imaging observations in connection with in situ measurements and find that the CME was a complex event composed of two eruptions separated by 10-15 min. They suggest that the extremely enhanced ejecta magnetic field and unusually high speed at STEREO A were caused by the in-transit interaction between the two closely launched eruptions and preconditioning of the upstream solar wind due to an earlier CME, respectively. \citet{temmer15} and \citet{cash15} investigate the propagation behavior of the 2012 July 23 CME/shock, both of which give results consistent with the preconditioning of the interplanetary medium by a previous CME proposed by \citet{liu14}. Another detailed analysis of the in situ shock properties indicates a quasi-parallel shock at STEREO A \citep{riley16}, again supporting the ``preconditioning" idea \citep{liu14}. The longitudinal distribution of the solar energetic particle event associated with the CME is studied by \citet{zhu16} using the particle data from STEREO and near-Earth spacecraft. 

Despite these previous studies of the 2012 July 23 complex CME, there are still plenty of data that have not been explored, such as: (1) the heliospheric imaging observations that allow to investigate how the CME/shock propagated and expanded in interplanetary space; (2) a radio type II burst (which we define here as a complex type II burst) that provides more information on the shock propagation and expansion; and (3) in situ measurements at different locations that may show the impact of the CME/shock or not. In this paper we use these data sets to address critical questions concerning CME-driven shocks: how does a CME-driven shock (in particular a large one) expand and propagate in the corona and interplanetary space? how extended can it be? and how are its spatial/temporal variations related to type II burst characteristics? We also present a new technique applicable to both coronagraph and heliospheric imaging observations for shock parameter determination. The paper is organized as follows. We describe multi-point imaging observations and modeling in Section 2, the characteristics of the associated complex type II burst in Section 3, and in situ measurements at different locations and MHD propagation beyond 1 AU in Section 4. The results are summarized and discussed in Section 5. These efforts provide insights into CME-driven shocks and have important implications for particle acceleration and space weather. 

\section{Multi-Point Imaging Observations and Modeling} 

Three widely separated spacecraft having imaging capabilities look at the Sun simultaneously, including the Solar and Heliospheric Observatory \citep[SOHO;][]{domingo95} at L1 and the two STEREO spacecraft (see Figure~1). On 2012 July 23, STEREO A and B were 0.96 AU and 1.02 AU from the Sun and 121.2$^{\circ}$ west and 114.8$^{\circ}$ east of the Earth (SOHO), respectively. The event originated from AR 11520 (S15$^{\circ}$W133$^{\circ}$), which was about 12$^{\circ}$ west of the central meridian as viewed from STEREO A. The Dawn spacecraft and Jupiter (not shown in Figure~1) were 2.54 AU and 5.02 AU from the Sun and 16.4$^{\circ}$ and 2$^{\circ}$ east of STEREO A, respectively. As they were almost radially aligned with STEREO A, the CME/shock may also have detectable signatures at these two locations. Venus or the Venus Express spacecraft around Venus, which was 29.8$^{\circ}$ west of the Earth and 0.73 AU from the Sun, and Mercury or the MESSENGER spacecraft orbiting Mercury, which was about 10$^{\circ}$ east of the Earth and about 0.45 AU from the Sun, are useful to test if the shock arrived there or not.          

Figure~2 shows the coronagraph observations and a spherical modeling of the shock. The shock, a faint edge ahead of the CME front, was already well developed and appeared nearly spherical in all the three views. LASCO C3 aboard SOHO even saw backward expansion of the shock on the opposite side of the Sun. For simplicity, we assume a spherical structure to model the shock (routines available under SSW), with which the expansion speed can be readily separated from the propagation speed of the shock sphere. This leads to four free parameters in the model, the longitude and latitude of the propagation direction and the distance ($d$) and radius ($r$) of the sphere. Reasonable fits are generally obtained by adjusting the model parameters to visually match the coronagraph images from STEREO A, B and SOHO simultaneously (see the bottom panels of Figure~2). Note that the time for the LASCO image shown in Figure~2 (03:30 UT) is about 6 min later than those for the STEREO images. The longitude of the propagation direction resulting from this forward modeling exhibits a change from about 16$^{\circ}$ to about 8$^{\circ}$ west of STEREO A, which roughly agrees with the eastward transition found by \citet{liu14}. The final propagation angle in the coronagraph images seems around 8$^{\circ}$ west of STEREO A and about 5$^{\circ}$ north in latitude.   

Composite images of COR2 and HI1 are shown in Figure~3, illustrating how the expansion of the CME/shock made it visible to HI1. COR2 has a field of view (FOV) of 0.7$^{\circ}$ - 4$^{\circ}$ around the Sun, and HI1 has a sided FOV of 4$^{\circ}$ - 24$^{\circ}$ along the Sun-Earth line (see Figure~1). A large lateral expansion is evident in the images from STEREO A: the CME appeared as a halo event in COR2, and soon a broad wave was seen in HI1, which is probably the shock and sheath dominating in the images. STEREO B mainly observed the backward expansion of the shock: the CME was headed towards the east in COR2, but backward expanding structures were discernible and finally made it to HI1, which is similar to what LASCO saw (see Figure~2). Note that the backward expanding feature must be the shock wave, not the ejected CME material which was traveling in the opposite direction. These scenarios as well as the locations of Mercury and Venus in the images are consistent with the geometries of the shock, the spacecraft and their FOVs as shown Figure~1. In particular, HI1 of STEREO B was in a good position to observe the backward expansion of the shock.     

Another way to show the expansion extent of the CME/shock is to produce time-elongation maps by stacking the running difference images within a slit along the ecliptic \citep[e.g.,][]{sheeley08, davies09, liu10a}, as displayed in Figure~4. The track in STEREO A, which represents the lateral expansion of the CME/shock, seems to reach the elongation of Mercury (about 14.4$^{\circ}$) although there was a data gap in HI1 on July 23. Note that this is an apparent approach to Mercury, i.e., not real (see the geometries in Figure~1). The track in STEREO B, which corresponds to the backward expansion of the shock, fails to reach the elongation of Mercury (about 21.1$^{\circ}$). The configuration in Figure~1 (specifically, the FOV of HI1 on STEREO B and the location of Mercury relative to the CME propagation direction) suggests that the shock was not able to reach Mercury, i.e., decayed in interplanetary space.

These imaging observations from three widely separated spacecraft indicate that the shock was expanding in all directions and eventually enclosed the whole Sun. This implies a shock geometry with an angular size of 360$^{\circ}$ and a moving center. Current geometries used to convert elongation to distance have either a limited angular width or a fixed center \citep[e.g.,][]{lugaz09, liu10b, davies13, mostl14}\footnote{It comes to our attention that the model 2 of \citet{lugaz10} might be applicable to the angular size of 360$^{\circ}$. That model also assumes tangents to a spherical structure but has been applied exclusively to CMEs with angular widths $\leqslant$180$^{\circ}$. Note that only CME-driven shocks/waves can have an angular size of 360$^{\circ}$, not the CME material.}. Here we present a new approach, which is applicable to both coronagraph and heliospheric imaging observations, to determine the shock propagation and expansion distances. It is similar to the triangulation technique originally proposed by \citet{liu10a, liu10b} and can provide results that may be compared with the above image forward modeling. The diagram shown in Figure~1 yields 
\begin{equation}
\frac{r}{\sin\alpha_{\rm A}} + d = d_{\rm A},
\end{equation}
\begin{equation}
d^2 + r^2 - 2dr\sin(\gamma-\alpha_{\rm B}) = d^2_{\rm B}\sin^2\alpha_{\rm B} + d^2\cos^2(\gamma-\alpha_{\rm B}). 
\end{equation}
The second equation is derived by combining $\epsilon = \gamma - \alpha_{\rm B} - \pi/2$ and $d^2 + r^2 - 2dr\cos\epsilon = d^2_{\rm B}\sin^2\alpha_{\rm B} + d^2\sin^2\epsilon$. Here we assume that the shock center has a fixed propagation direction towards STEREO A. The 8$^{\circ}$ deviation from the longitude of STEREO A resulting from coronagraph image forward modeling is considerably small and may not cause significant errors in the results (the error can be estimated by $\Delta d/d \sim \tan8^\circ = 14\%$). Our previous study gives an average deviation of about 2$^{\circ}$ from the direction of STEREO A \citep{liu14}, and in situ analysis of the shock near 1 AU indicates that the shock nose in the ecliptic is very close to STEREO A \citep{riley16}. The elongation angles ($\alpha_{\rm A}$ and $\alpha_{\rm B}$) can be obtained from the tracks in the time-elongation maps (Figure~4). The two equations can then be solved for $r$ and $d$. There are actually two solutions with $d<r$ and $d>r$, respectively. Only the solution of $d<r$ is valid for the current case, as otherwise no backward expansion on the opposite side of the Sun would be observed. Note that we apply this method only to the COR2 and HI1 data which show the backward expansion.

Figure~5 shows the comparison of the radial distance and speed of the CME/shock nose between different methods. For the image forward modeling and the present triangulation with STEREO A and B, the ``nominal" distance of the CME/shock nose is $r+d$ by definition. The image forward modeling gives distances and speeds generally consistent with but slightly lower than those from the SOHO and STEREO B triangulation \citep{liu14}. This difference is reasonable and likely owing to the fact that the forward modeling deals with the whole shock structure (see Figure~2) while the SOHO and STEREO B triangulation of \citet{liu14} is intended to look at only the nose in the ecliptic. The new triangulation yields even lower distances and speeds for the CME/shock nose, but roughly connects with the forward modeling around 04-05 UT. The speed from the new triangulation quickly decreases to about 1300 km s$^{-1}$, which is unlikely to be true for the nose. Note that the new triangulation is designed for the lateral/backward expansion of the shock, not the motion of the nose (see Figure~1). If the shock were perfectly spherical, the nose speeds determined from the three methods would be comparable. This comparison implies that at later times the shock nose is faster than what is inferred from a spherical morphology, so the shock structure around the nose is non-spherical at large distances, probably more like a planetary bow shock.

Then we examine the propagation and expansion of the shock separately, as plotted in Figure~6. Again, the results from the new triangulation roughly match those from the image forward modeling around 04-05 UT. The radius of the shock sphere ($r$) is generally larger than the distance of the center ($d$) except for the first two data points, so the expansion of the shock largely dominates over the propagation for the present time period. Their ratio first increases quickly and then flattens. If this ratio is taken as a measure of self-similarity, then self-similar expansion is roughly valid only after the first $\sim$1.5 hours. The expansion speed has a maximum value of about 1500 km s$^{-1}$ and keeps larger than the propagation speed by about 400 km s$^{-1}$ during the time period. Both the expansion and propagation slow down at later times, as indicated by a decrease in both of the speeds. Again, this should not be taken as a sign of the slowdown of the CME/shock nose, as the data used for the new triangulation represent more of the lateral/backward expansion (see Figure~1).

\section{Complex Radio Type II Burst} 

Figure~7 displays the radio dynamic spectra associated with the event. In STEREO A observations type II emissions appear as many drifting bands and cover a large range of frequencies, from the upper bound (16 MHz) intermittently down to about 15 kHz. There are two bands, although not continuous, that were observed by all the three spacecraft (marked with diamonds). These two bands can be interpreted as a fundamental-harmonic (F-H) pair. If we use the SOHO and STEREO B triangulation distances \citep{liu14} and a Leblanc density model \citep{leblanc98} to simulate the two bands, an ambient solar wind density of $n_0=30$ cm$^{-3}$ at 1 AU would be required. The frequency-time curves corresponding to this F-H pair (diamonds) capture many type II features extending from high to low frequencies, including a spot at about 150 kHz observed by both STEREO A and B around 13 UT. The F-H pair likely resulted from a shock component propagating into a high-density region. There are also some low-frequency type II features observed by only STEREO A especially after 12 UT, implying a very low ambient solar wind density. Some of these type II bands can indeed be explained by an F-H pair based on the SOHO and STEREO B triangulation distances \citep{liu14} and a $n_0=1.5$ cm$^{-3}$ Leblanc density model (marked with crosses). This F-H pair was probably caused by a shock component moving into a low-density region towards STEREO A. This result agrees very well with the in situ measurements of a quite low density ($\sim$1 cm$^{-3}$) ahead of the shock at STEREO A, which is interpreted as a consequence of preconditioning of the interplanetary medium by an earlier CME \citep{liu14}.  

Based on the above observations and analysis, here we define a complex type II burst as follows: a type II burst composed of multiple branches that are not all F-H related. The drift rates of the branches are not necessarily the same. In the current case, we see at least two F-H pairs (not even mentioning the two bands above $\sim$300 kHz observed by STEREO A between 04-05 UT). The high-density F-H pair (marked with diamonds) were observed by all the three spacecraft before 06 UT, but the low-density bands (marked with crosses) were detected only by STEREO A in particular after 12 UT. These two pairs seem to have come from the same shock, but different parts of the shock that were propagating into different regions with dissimilar densities. At earlier times the shock may have a larger angular extent or stronger emissions observable to all the three spacecraft, whereas at later times the major emitting component was only towards STEREO A. These results reveal how a single shock can produce diverse, complex radio emission characteristics, depending on the motion and evolution of its different components and where these components travel. 

The spectrum indicates shock arrival at STEREO A and B, as can be seen from the suddenly enhanced, diffuse intensity. The shock arrival times revealed by the spectrum are consistent with the in situ measurements at the two spacecraft (see Figures~8 and 9). As will be explained later, the shock at STEREO B is different from that at STEREO A. It is not clear whether the shock observed at STEREO B had generated any type II burst, but our analysis of the spectra for July 23 suggests that all the type II bands may have originated from the shock observed at STEREO A. Also shown in Figure 7 is the normalized full-disk EUV flux at 195 \AA, which has two maxima. The first maximum arose from the flare radiation associated with the two consecutive prominence eruptions \citep{liu14}, while the second one, which is even higher, wider and coincident with the shock arrival at STEREO A, was due to the impact of energetic particles on the EUV detector. After the shock, the EUV flux shows a two-step decrease, first in the sheath between the shock and ejecta and then in the first ICME \citep{liu14}. The second reduction is probably owing to exclusion of the energetic particles by the strong magnetic field inside the ejecta \citep[e.g.,][]{liu08}. The two-step decrease is similar to the behavior of the intensity of $>$10 MeV particles across the shock, sheath and the first ICME \citep{zhu16}. This similarity may indicate that mainly the $>$10 MeV particles resulted in the CCD response for the current case. These results illustrate how EUV, radio and in situ measurements including energetic particles all give a consistent story on shock arrival.   

\section{Multi-Point In Situ Measurements and MHD Propagation} 

The in situ signatures at STEREO A (0.96 AU) are shown in Figure~8, which indicate a fast, forward shock followed by two ICMEs. The readers are directed to \citet{liu14} for a detailed discussion of the in situ signatures. Here we briefly summarize some of the results relevant to the present work. The shock passed STEREO A at 20:55 UT on July 23, only about 18.6 hours after the CME launch time on the Sun ($\sim$02:20 UT). The shock normal is predominantly along the radial direction from the Sun \citep{riley16}, so STEREO A should be very close to the shock nose in the ecliptic. Reconstruction using a Grad-Shafranov (GS) technique \citep{hau99, hu02} gives a right-handed structure for both of the ICMEs and very oblique axis orientations \citep{liu14}.    

Figure~9 shows the in situ signatures at STEREO B (1.02 AU). A shock passed STEREO B at 21:22 UT, only 27 min later than the shock arrival at STEREO A. An ICME with enhanced magnetic fields and clear rotation of the fields started about 20.6 hours after the shock. The GS reconstruction yields a well-organized, left-handed flux-rope structure for the ICME with an axis elevation angle of about $-48^{\circ}$ and azimuthal angle of about $93^{\circ}$ in RTN coordinates (not shown here). The handedness is contrary to those of the ICMEs at STEREO A, so this is a different ICME. The sheath size (20.6 hours) is unusually long compared with the average value of 14 hours at 1 AU \citep{liu06b}. Also note that the speed in the sheath first decreased and then rose as well as the complex variations of the sheath field components. These unusual characteristics of the sheath raise a doubt if the shock was driven by the following ICME. 

If the shock were associated with the July 23 event, this would imply an average transit speed of about 2200 km s$^{-1}$ from the Sun to STEREO B. However, the maximum expansion speed of the shock is 1500 km s$^{-1}$ (Figure~6), not even to mention the offset by the translation of the shock center towards STEREO A (Figure~1). Therefore, it is unlikely the same shock as observed at STEREO A. The speed across the shock and ICME is generally below 400 km s$^{-1}$, which suggests that the solar source eruption(s) occurred probably several days before July 23. The data after the ICME do not show a shock until 14:59 UT on July 28 (not shown here), which seems too late to be associated with the July 23 event. The speed downstream of the shock is about 450 km s$^{-1}$, which indicates a solar source release time no earlier than mid July 24. The shock driven by the July 23 complex CME probably faded before reaching STEREO B in this lateral direction.   

We have also examined magnetic field measurements by MESSENGER at 0.45 AU from the Sun and by Venus Express at 0.73 AU as well as the solar wind plasma and magnetic field data from Wind near the Earth (not shown here). No shock can be unambiguously associated with the July 23 event from the in situ measurements at these three locations. Note that any shock arrival at Mercury, Venus and the Earth due to the July 23 event has to be consistent with the timing based on the backward/lateral velocity of the shock. The backward speed of the shock, the difference between the expansion and translation speeds, is at most $\sim$500 km s$^{-1}$, and the lateral speed is no more than 1500 km s$^{-1}$ (see Figures~1 and 6). The in situ results agree with our image analysis that indicates the decay of the shock before reaching the distance of Mercury in the backward direction (see Section 2). One may suspect that the shock could be too faint to be revealed by the heliospheric images and time-elongation map of STEREO B. We extrapolate the track in the time-elongation map of STEREO B (Figure~4) out to Mercury's elongation, but do not find a shock signature around the intersection time from the in situ measurements at MESSENGER.

As discussed earlier, the July 23 event may cause interplanetary disturbances detectable at Dawn and Jupiter. We propagate the solar wind data outward from STEREO A using an MHD model, in order to look at the evolution beyond 1 AU (Figure~10). The model assumes spherical symmetry \citep{wang00} and has successfully connected solar wind measurements at different spacecraft \citep[e.g.,][]{wang01, liu06c, richardson06}. Hourly averages of the solar wind measurements at STEREO A are used as input to the model, which smoothes the speed downstream of the shock to about 2000 km s$^{-1}$ (top panel of Figure~10). The predicted arrival time of the shock is about 06 UT on July 25 at Dawn and about 23:30 UT on July 27 at Jupiter. The transit time from STEREO A to Jupiter is only about 4 days. We leave these predictions here, as they may be useful for future possible comparison with in situ measurements at Dawn and Jupiter which we cannot get for this study. The peak solar wind speed is still very high even at Jupiter (about 1550 km s$^{-1}$), so the shock would persist out to much larger distances around the nose direction. 

\section{Conclusions and Discussion}

We have investigated the structure, propagation and expansion of the shock associated with the 2012 July 23 extreme storm from the Sun far into interplanetary space, combining wide-angle heliospheric imaging observations, a complex radio type II burst and in situ measurements from a fleet of spacecraft. Although the focus is on what we call an ``extreme" event, shocks driven by wide, fast CMEs may share similar characteristics as derived here. The results are summarized and discussed as follows. 

1. The shock was expanding in all directions and appeared nearly spherical as viewed from three widely separated spacecraft. STEREO B and SOHO observed even backward expansion of the shock on the side of the Sun opposite to the CME propagation direction. The July 23 complex CME relatively separated from other eruptions, so the observed nearly Sun-surrounding wave is unlikely a result of connecting with other events. The shock can be modeled reasonably well by a spherical structure in both coronagraph and heliospheric images. The large angular extent of the shock suggests that a wide area of the inner heliosphere can be disturbed. It also implies that energetic particles can have a vast source region of energization and thus a wide distribution in both longitude and latitude \citep[e.g.,][]{liu11, rouillard12, lario14}. Widespread energetic particles were indeed observed at 1 AU associated with the July 23 event \citep{zhu16}, although there could be contributions from other eruptions. 

2. Our separation of expansion from translation suggests that the expansion of the shock largely dominates over its propagation for the time period we have studied. The expansion is roughly self-similar, except the first $\sim$1.5 hours from launch. The expansion speed of the shock is as high as 1500 km s$^{-1}$ and keeps larger than the propagation speed of the shock center by about 400 km s$^{-1}$. (Note that the center of the shock should not be taken as the center of the ejecta.) This indicates an enormous internal pressure of the ejecta that drove the expansion at early times. However, the expansion due to the internal pressure is expected to decrease quickly at later times, as we frequently see magnetic clouds at 1 AU with low temperatures and nearly force-free magnetic field configurations \citep[see more discussions in][]{liu06a}. For a force-free configuration the magnetic pressure is balanced by the field line tension. 

3. Our study also gives indications of a bow-shock morphology around the nose at later times. The ``nominal" nose velocity from a new triangulation method, which we present here for the study of shock expansion, quickly decreases in contrast with the roughly constant speed from the SOHO and STEREO B triangulation \citep{liu14}. This contrast implies that the shock structure around the nose is non-spherical especially at large distances. The shock is probably more like a planetary bow shock around the nose due to the outward motion in combination with the expansion of the ejecta. At other parts the shock is generated predominantly by the lateral/backward expansion of the ejecta. In any case the shock is piston driven, with the piston being the ejecta \citep{liu09a}. It is worth noting that both planetary bow shocks and CME-driven shocks have the same basic physics: a shock forms when the supersonic, super-Alfv\'{e}nic solar wind encounters an obstacle. However, CMEs are expanding obstacles while planetary magnetospheres are non-expanding ones \citep[see discussions in][]{liu06b, richardson07, siscoe08}. 

4. The results suggest fading and persistence of the shock along different directions in interplanetary space. When the influence of the expansion subsides depending on the distance from the Sun and/or internal pressure of the ejecta, the shock may decay in particular at the wake. Heliospheric images from STEREO B indicate that the shock in the backward direction decayed before reaching the distance of Mercury. Coordinated in situ measurements from MESSENGER and Wind support this interpretation. It is also unlikely that the shock sustained out to STEREO B and Venus in the lateral directions. Our MHD propagation of the solar wind disturbance observed at STEREO A, however, shows that the shock in the nose direction may persist to the far outer heliosphere, with possible impact on Dawn around 06 UT on July 25 and on Jupiter around 23:30 UT on July 27. These results imply that the term ``shock", which we have used throughout the paper, may not apply all the time. At some point especially for the wake, the structure in question could be just a wave that had lost its non-linear steepening character.

5. Characteristics of the associated complex type II burst are consistent with the spatial and temporal variations of the shock deduced from imaging and in situ observations. A complex type II burst is composed of multiple branches that are not all F-H related, as defined here. We observe at least two separate F-H pairs, produced by different parts of the shock that were propagating into regions with quite different densities. The high-density F-H pair were observed by all the three spacecraft before 06 UT on July 23, suggestive of a larger angular extent of the shock or stronger emissions observable to all the three spacecraft at earlier times. The low-density F-H bands were detected only by STEREO A especially after 12 UT on July 23, which indicates that the major emitting component was only towards STEREO A at later times. This agrees remarkably well with the in situ measurements of a very low density ($\sim$1 cm$^{-3}$) ahead of the shock at STEREO A due to the preconditioning by an earlier CME \citep{liu14}. 

\acknowledgments The research was supported by the Recruitment Program of Global Experts of China, NSFC under grant 41374173 and the Specialized Research Fund for State Key Laboratories of China. We acknowledge the use of data from STEREO, SOHO, Wind, MESSENGER and Venus Express.

\clearpage

\begin{figure}
\epsscale{0.7} \plotone{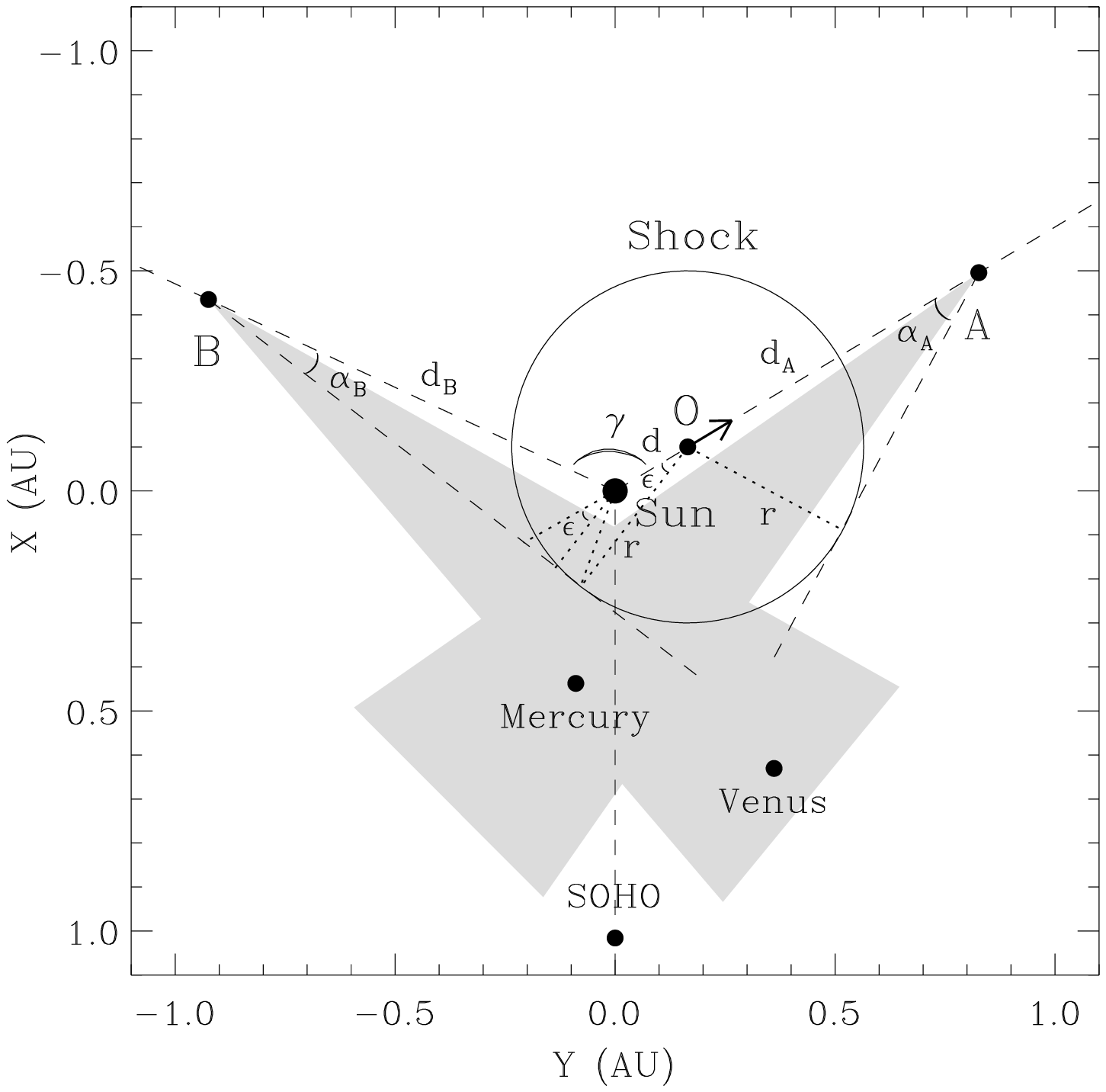} 
\caption{Positions of the spacecraft and planets on 2012 July 23. Also shown is a diagram used to determine the propagation and expansion of the CME-driven shock in the ecliptic plane. The grey areas are the fields of view of HI1 on STEREO A and B, respectively. The shock is assumed as a spherical structure propagating towards STEREO A. The distance of its center (O) from the Sun is denoted as $d$ and its radius as $r$. Elongation angles ($\alpha_{\rm A}$ and $\alpha_{\rm B}$) of the shock are measured along its tangents by STEREO A and B. Also indicated are the distances of the two spacecraft from the Sun ($d_{\rm A}$ and $d_{\rm B}$) and their longitudinal separation ($\gamma$) behind the Sun. The dotted lines and the angle $\epsilon$ help illustrate how Equations (1) and (2) are derived.}
\end{figure}

\clearpage

\begin{figure}
\epsscale{0.9} \plotone{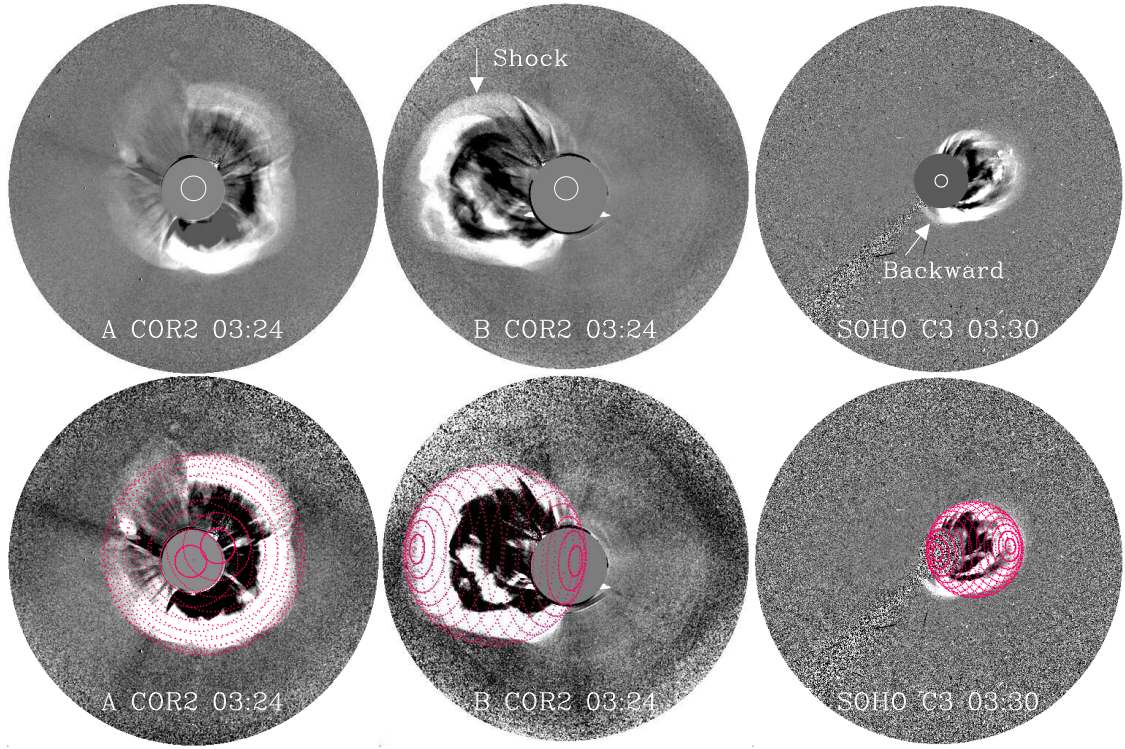} 
\caption{Top: running-difference coronagraph images of the CME and shock from STEREO A (left), B (middle) and SOHO (right). Note the backward expansion of the CME-driven shock in LASCO C3 of SOHO. Bottom: spherical wireframe rendering of the CME-driven shock superposed on the observed images. A sharper contrast is used in the running differences to enhance the visibility of the shock.}
\end{figure}

\clearpage

\begin{figure}
\epsscale{1.0} \plotone{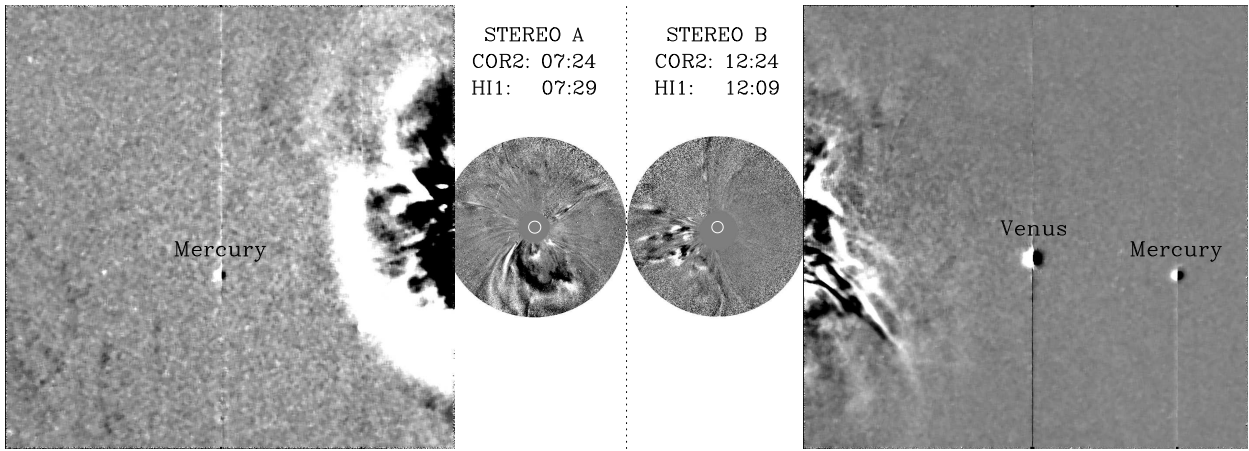} 
\caption{Composite running-difference images of COR2 and HI1 aboard STEREO A (left) and B (right). The positions of Venus and Mercury are labeled in the HI1 images. Animations associated with this figure are available in the online journal, showing how the expansion of the CME/shock made the event visible in HI1.}
\end{figure}

\clearpage

\begin{figure}
\epsscale{0.7} \plotone{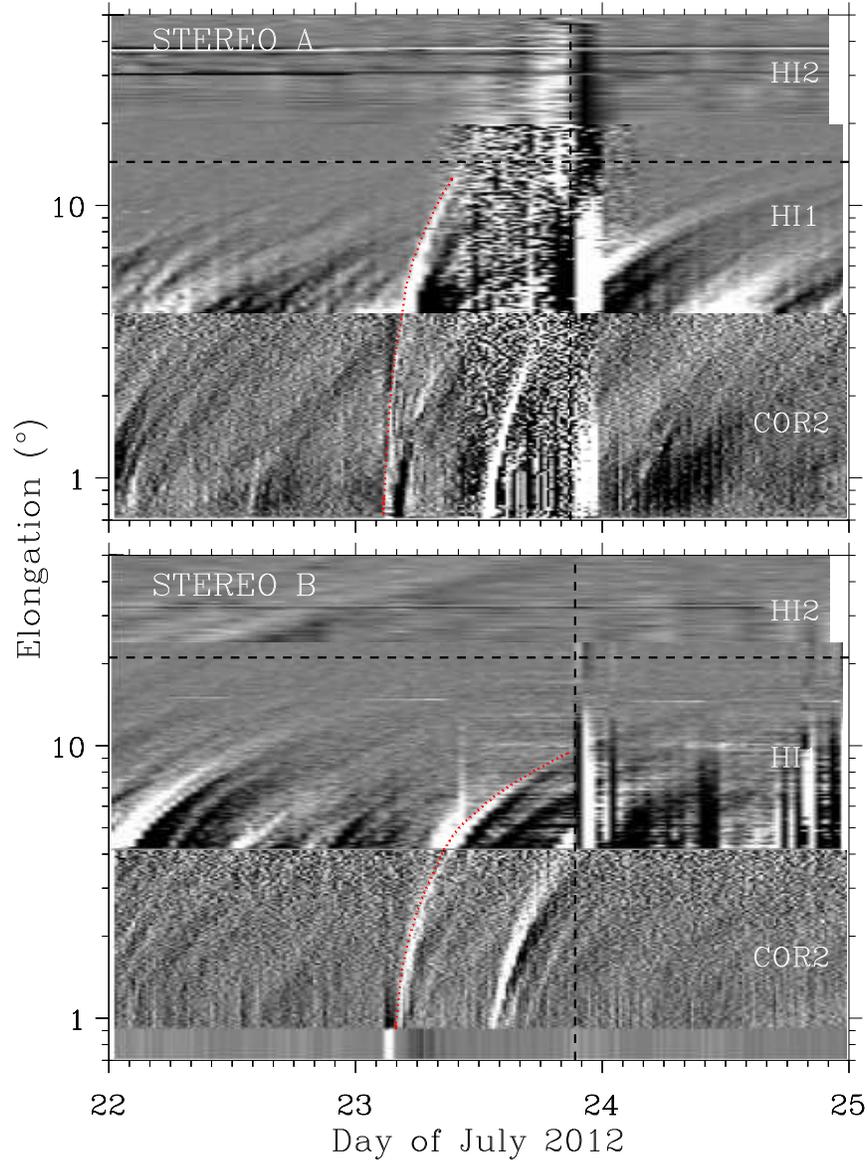} 
\caption{Time-elongation maps constructed from running-difference images of COR2, HI1 and HI2 along the ecliptic. Note that, to connect with HI observations, we use data from the eastern part of COR2 for STEREO A ($90^{\circ}$ counterclockwise from the ecliptic north) and data from the western part of COR2 for STEREO B ($90^{\circ}$ clockwise from the ecliptic north). The red dotted curve indicates the track of the event, along which the elongation angles are extracted. The vertical dashed line marks the observed shock arrival time at each spacecraft. The horizontal dashed line denotes the elongation angle of Mercury.}
\end{figure}

\clearpage

\begin{figure}
\epsscale{0.7} \plotone{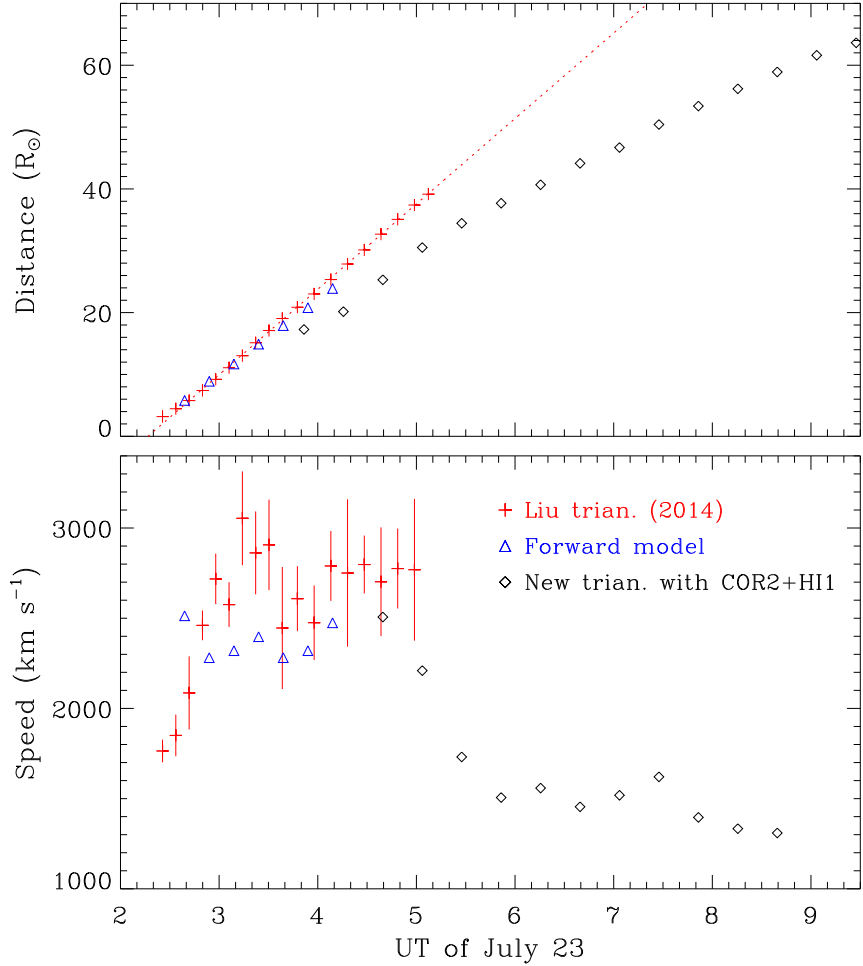} 
\caption{Radial distance and speed of the CME/shock nose. The speeds are calculated from adjacent distances with a numerical differentiation technique. The red crosses represent the results obtained by \citet{liu14} using SOHO and STEREO B triangulation, the blue triangles show the outcome from image forward modeling (see Figure~2), and the black diamonds are those from the present triangulation with STEREO A and B (see Figure~1). The image forward modeling and the new triangulation cover different time periods, but their results roughly match around 04-05 UT. The red dotted line is a linear fit of the red-cross distances with an average speed of about 2670 km s$^{-1}$.}
\end{figure}

\clearpage

\begin{figure}
\epsscale{0.7} \plotone{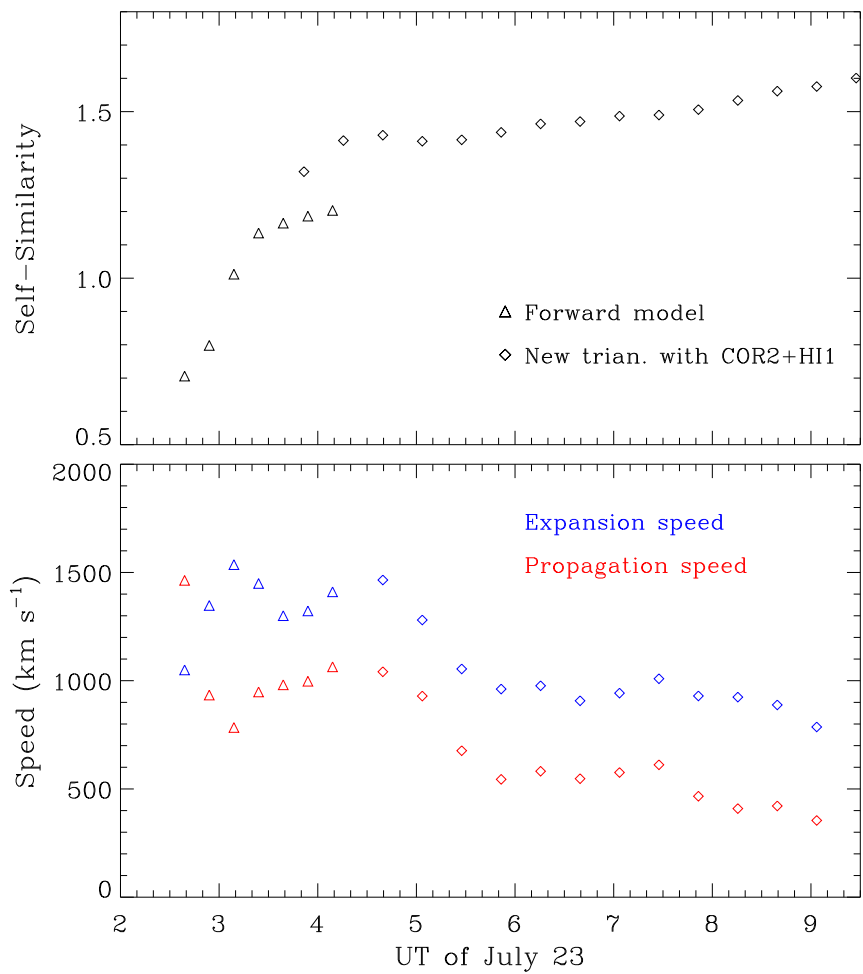} 
\caption{Self-similarity (defined as $r/d$) and propagation (red) and expansion (blue) speeds of the CME-driven shock. The propagation speed is derived from the numerical differentiation of the shock center distance $d$, while the expansion speed is calculated from the radius $r$ of the spherical shock. The triangles and diamonds represent results from image forward modeling (Figure~2) and triangulation with STEREO A and B (Figure~1), respectively.}
\end{figure}

\clearpage

\begin{figure}
\epsscale{0.7} \plotone{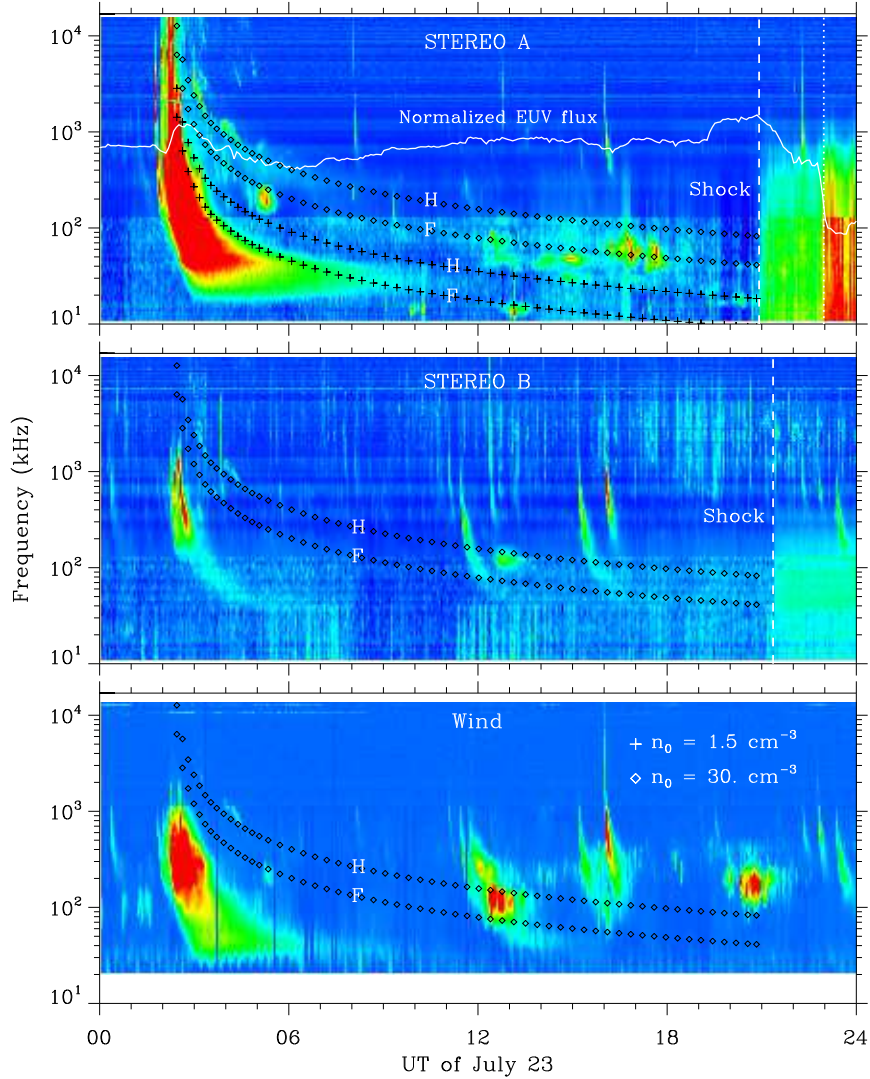} 
\caption{Dynamic spectrum from STEREO A, B and Wind. A complex type II burst is observed with multiple branches. The distances from SOHO and STEREO B triangulation \citep{liu14} and their linear extrapolation (see Figure~5) are converted to frequencies by using the Leblanc density model with a 1-AU density of 1.5 (crosses) and 30 cm$^{-3}$ (diamonds), respectively. The resulting fundamental (F) and second harmonic (H) frequencies are then plotted over the dynamic spectrum. The white curve in the upper panel is the normalized full-disk EUV flux at 195 \AA\ observed by STEREO A. The vertical dashed line marks the shock arrival time from in situ measurements at each STEREO spacecraft. The vertical dotted line in the upper panel indicates the leading boundary of the first ICME \citep{liu14}.}
\end{figure}

\clearpage

\begin{figure}
\epsscale{0.7} \plotone{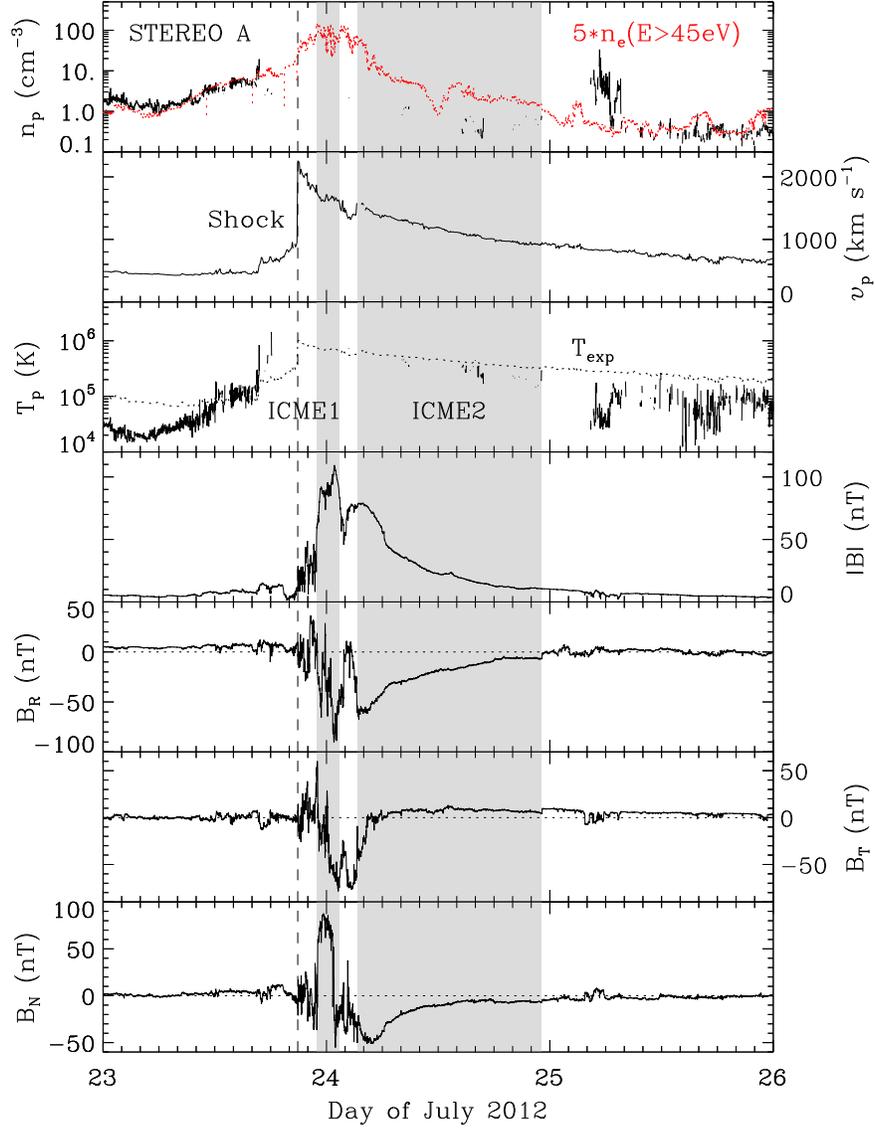} 
\caption{Solar wind measurements at STEREO A \citep[adapted from][]{liu14}. From top to bottom, the panels show the proton density, bulk speed, proton temperature, and magnetic field strength and components, respectively. The shaded regions indicate the ICME intervals, and the vertical dashed line marks the shock arrival. The red curve in the top panel represents the number density (multiplied by a factor of 5) of electrons with energies above 45 eV. The dotted curve in the third panel denotes the expected proton temperature calculated from the observed speed \citep{lopez87}.}
\end{figure}

\clearpage

\begin{figure}
\epsscale{0.7} \plotone{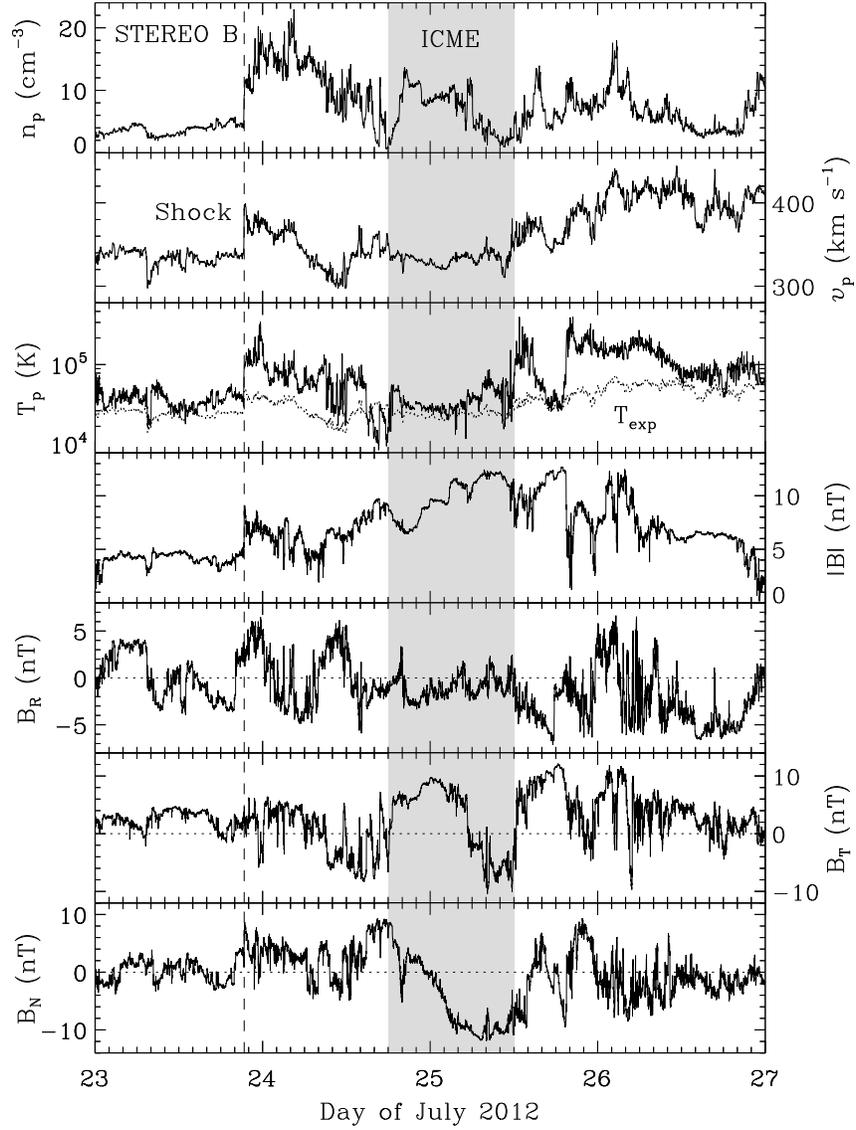} 
\caption{Solar wind measurements at STEREO B. Similar to Figure~8. A single ICME with an usually long sheath is observed during the time period.}
\end{figure}

\clearpage

\begin{figure}
\epsscale{0.7} \plotone{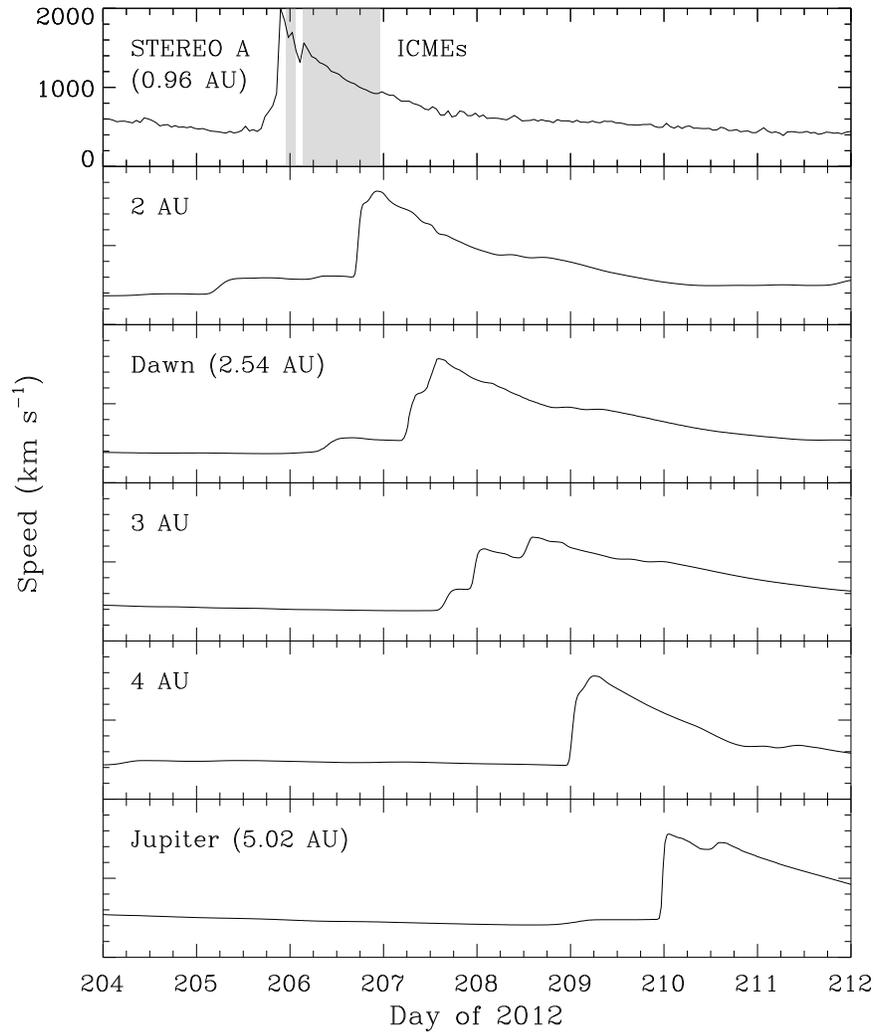} 
\caption{Evolution of solar wind speeds from STEREO A to Jupiter via the MHD model. The curve in the top panel is the observed speed at STEREO A, and others are predicted speeds at given distances. The shaded regions represent the observed ICME intervals.}
\end{figure}


\begin{thebibliography}{}

\bibitem[Baker et al.(2013)]{baker13}
Baker, D. N., Li, X., Pulkkinen, A., et al. 2013, SpWea, 11, 585

\bibitem[Cane et al.(1987)]{cane87}
Cane, H. V., Sheeley Jr., N. R., \& Howard, R. A. 1987, JGR, 92, 9869

\bibitem[Cash et al.(2015)]{cash15}
Cash, M. D., Biesecker, D. A., Pizzo, V., et al. 2015, SpWea, 13, 611 

\bibitem[Davies et al.(2009)]{davies09}
Davies, J. A., Harrison, R. A., Rouillard, A. P., et al. 2009, GeoRL, 36, L02102

\bibitem[Davies et al.(2013)]{davies13}
Davies, J. A., Perry, C. H., Trines, R. M. G. M, et al. 2013, ApJ, 777, 167

\bibitem[Domingo et al.(1995)]{domingo95}
Domingo, V., Fleck, B., \& Poland, A. I. 1995, SoPh, 162, 1  

\bibitem[Feng et al.(2012)]{feng12}
Feng, S. W., Chen, Y., Kong, X. L., et al. 2012, ApJ, 753, 21

\bibitem[Hau \& Sonnerup(1999)]{hau99}
Hau, L.-N., \& Sonnerup, B. U. \"{O}. 1999, JGR, 104, 6899

\bibitem[Hess \& Zhang(2014)]{hess14}
Hess, P., \& Zhang, J. 2014, ApJ, 792, 49

\bibitem[Hu et al.(2016)]{hu16}
Hu, H., Liu, Y. D., Wang, R., M\"{o}stl, C., \& Yang, Z. 2016, ApJ, in press 

\bibitem[Hu \& Sonnerup(2002)]{hu02}
Hu, Q., \& Sonnerup, B. U. \"{O}. 2002, JGR, 107, 1142

\bibitem[Kaiser et al.(2008)]{kaiser08}
Kaiser, M. L., Kucera, T. A., Davila, J. M., et al. 2008, SSRv, 136, 5

\bibitem[Kwon et al.(2014)]{kwon14}
Kwon, R., Zhang, J., \& Olmedo, O. 2014, ApJ, 794, 148

\bibitem[Kwon et al.(2015)]{kwon15}
Kwon, R., Zhang, J., \& Vourlidas, A. 2015, ApJL, 799, L29

\bibitem[Lario et al.(2014)]{lario14}
Lario, D., Raouafi, N. E., Kwon, R.-Y., et al., 2014, ApJ, 797, 8

\bibitem[Leblanc et al.(1998)]{leblanc98}
Leblanc, Y., Dulk, G. A., \& Bougeret, J.-L. 1998, SoPh, 183, 165

\bibitem[Liu et al.(2006a)]{liu06a}
Liu, Y., Richardson, J. D., Belcher, J. W., Kasper, J. C., \& Elliott, H. A. 2006a, JGR, 111, A01102 

\bibitem[Liu et al.(2006b)]{liu06b}
Liu, Y., Richardson, J. D., Belcher, J. W., Kasper, J. C., \& Skoug, R. M. 2006b, JGR, 111, A09108 

\bibitem[Liu et al.(2006c)]{liu06c}
Liu, Y., Richardson, J. D., Belcher, J. W., Wang, C., Hu, Q., \& Kasper, J. C. 2006c, JGR, 111, A12S03

\bibitem[Liu et al.(2008)]{liu08}
Liu, Y., Luhmann, J. G., M\"{u}ller-Mellin, R., et al. 2008, ApJ, 689, 563

\bibitem[Liu et al.(2009a)]{liu09a}
Liu, Y., Luhmann, J. G., Bale, S. D., \& Lin, R. P. 2009a, ApJL, 691, L151 

\bibitem[Liu et al.(2009b)]{liu09b}
Liu, Y., Luhmann, J. G., Lin, R. P., et al. 2009b, ApJL, 698, L51

\bibitem[Liu et al.(2010a)]{liu10a}
Liu, Y., Davies, J. A., Luhmann, J. G., et al. 2010a, ApJL, 710, L82

\bibitem[Liu et al.(2010b)]{liu10b}
Liu, Y., Thernisien, A., Luhmann, J. G., et al. 2010b, ApJ, 722, 1762

\bibitem[Liu et al.(2011)]{liu11}
Liu, Y., Luhmann, J. G., Bale, S. D., \& Lin, R. P. 2011, ApJ, 734, 84

\bibitem[Liu et al.(2012)]{liu12}
Liu, Y. D., Luhmann, J. G., M\"{o}stl, C., et al. 2012, ApJL, 746, L15

\bibitem[Liu et al.(2013)]{liu13}
Liu, Y. D., Luhmann, J. G., Lugaz, N., et al. 2013, ApJ, 769, 45

\bibitem[Liu et al.(2014)]{liu14}
Liu, Y. D., Luhmann, J. G., Kajdi\v{c}, P., et al. 2014, NatCo, 5, 3481

\bibitem[Lopez(1987)]{lopez87}
Lopez, R. E. 1987, JGR, 92, 11189

\bibitem[Lugaz et al.(2009)]{lugaz09}
Lugaz, N., Vourlidas, A., \& Roussev, I. I. 2009, AnGeo, 27, 3479

\bibitem[Lugaz et al.(2010)]{lugaz10}
Lugaz, N., Hernandez-Charpak, J. N., Roussev, I. I., et al. 2010, ApJ, 715, 493

\bibitem[Maloney \& Gallagher(2011)]{maloney11}
Maloney, S. A., \& Gallagher, P. T. 2011, ApJL, 736, L5

\bibitem[Martinez-Oliveros et al.(2012)]{juan12}
Martinez-Oliveros, J. C., Raftery, C. L., Bain, H. M., et al. 2012, ApJ, 748, 66

\bibitem[M\"{o}stl et al.(2014)]{mostl14}
M\"{o}stl, C., Amla, K., Hall, J. R., et al. 2014, ApJ, 787, 119

\bibitem[Nelson \& Melrose(1985)]{nelson85}
Nelson, G. J., \& Melrose, D. B. 1985, in Solar Radiophysics:
Studies of Emission from the Sun at Metre Wavelengths, ed. D. J.
McLean \& N. R. Labrum (Cambridge: Cambridge Univ. Press), 333

\bibitem[Ngwira et al.(2013)]{ngwira13}
Ngwira, C. M., Pulkkinen, A., Leila Mays, M., et al. 2013, SpWea, 11, 671

\bibitem[Ontiveros \& Vourlidas(2009)]{ontiveros09}
Ontiveros, V., \& Vourlidas, A. 2009, ApJ, 693, 267

\bibitem[Reiner et al.(2007)]{reiner07}
Reiner, M. J., Kaiser, M. L., \& Bougeret, J.-L. 2007, ApJ, 663, 1369

\bibitem[Reisenfeld et al.(2003)]{reisenfeld03}
Reisenfeld, D. B., Gosling, J. T., Forsyth, R. J., Riley, P., \& St. Cyr, O. C. 2003, GeoRL, 30, 8031

\bibitem[Richardson et al.(2006)]{richardson06}
Richardson, J. D., Liu, Y., Wang, C., et al. 2006, GeoRL, 33, L23107

\bibitem[Richardson \& Liu(2007)]{richardson07}
Richardson, J. D., \& Liu, Y. 2007, in AIP Conf. Proc. 932, Turbulence and Nonlinear Processes in Astrophysical Plasmas, ed. D. Shaikh \& G. P. Zank (Melville, NY: AIP), 387

\bibitem[Riley et al.(2016)]{riley16}
Riley, P., Caplan, R. M., Giacalone, J., Lario, D., \& Liu, Y. 2016, ApJ, 819, 57

\bibitem[Rouillard et al.(2012)]{rouillard12}
Rouillard, A. P., Sheeley, N. R., Tylka, A., et al. 2012, ApJ, 752, 44 

\bibitem[Russell et al.(2013)]{russell13}
Russell, C. T., Mewaldt, R. A., Luhmann, J. G., et al. 2013, ApJ, 770, 38

\bibitem[Sheeley et al.(2008)]{sheeley08}
Sheeley, N. R., Herbst, A. D., Palatchi, C. A., et al. 2008, ApJ, 675, 853

\bibitem[Siscoe \& Odstrcil(2008)]{siscoe08}
Siscoe, G., \& Odstrcil, D. 2008, JGR, 113, A00B07  

\bibitem[Temmer \& Nitta(2015)]{temmer15}
Temmer, M., \& Nitta, N. V. 2015, SoPh, 290, 919	

\bibitem[Volpes \& Bothmer(2015)]{volpes15}
Volpes, L., \& Bothmer, V. 2015, SoPh, 290, 3005

\bibitem[Vourlidas et al.(2003)]{vourlidas03}
Vourlidas, A., Wu, S. T., Wang, A. H., Subramanian, P., \& Howard, R. A. 2003, ApJ, 598, 1392

\bibitem[Vourlidas \& Ontiveros(2009)]{vourlidas09}
Vourlidas, A., \& Ontiveros, V. 2009, AIP Conf. Proc., 1183, 139

\bibitem[Wang et al.(2000)]{wang00}
Wang, C., Richardson, J. D., \& Gosling, J. T. 2000, JGR, 105, 2337

\bibitem[Wang et al.(2001)]{wang01}
Wang, C., Richardson, J. D., \& Paularena, K. I. 2001, JGR, 106, 13007

\bibitem[Zhang et al.(2016)]{zhang16}
Zhang, J. J., Wang, C., Sun, T. R., \& Liu, Y. D. 2016, SpWea, 14, 259

\bibitem[Zhu et al.(2016)]{zhu16}
Zhu, B., Liu, Y. D., Luhmann, J. G., et al. 2016, ApJ, 827, 146

\end{thebibliography}
\end{document}